\documentclass[aps,twocolumn,superscriptaddress,amsfonts,amsmath,amssymb]{revtex4}
\usepackage{graphicx}
\usepackage{amssymb}
\usepackage{color}

\begin{document}

\title{Temporal self-restoration of compressed optical filaments}

\author{L. Berg\'e}
\affiliation{CEA-DIF, DPTA, Bruy\`eres-le-Ch\^atel, 91297 Arpajon
cedex, France}
\author{S. Skupin}
\affiliation{Max-Planck Institut f{\"u}r Physik Komplexer Systeme,
N{\"o}thnitzer Str. 38, 01187 Dresden, Germany}
\author{G. Steinmeyer}
\affiliation{Max-Born-Institut f\"ur Nichtlineare Optik und
Kurzzeitspektroskopie, Max-Born-Str. 2a, 12489 Berlin, Germany}

\begin{abstract}
We numerically investigate the propagation of a self-compressed optical 
filament
through a gas-glass-gas interface. Few-cycle light pulses survive a sudden 
and short
order-of-magnitude increase of nonlinearity and dispersion, even
when all conservative estimates predict temporal spreading or spatial breakup. 
Spatio-temporal distortions are shown to self-heal upon further
propagation when the pulse refocuses in the second gas.
This self-healing mechanism has important
implications for pulse compression techniques handled by filamentation and explains the robustness 
of such sources.
\end{abstract}

\maketitle

The physics of ultrashort optical filaments has recently received
increased attention as it enables self-compression of intense laser
pulses into the few-cycle regime \cite{Hauri:apb:79:673,
Stibenz:ol:31:274,Skupin:pre:74:056604,Champeaux:pre:68:066603,
Zair:oe:15:5394,Kosareva:apb:91:35, Berge:prl:100:113902}.
Femtosecond pulses with input powers close to the self-focusing
threshold, $P_{\rm cr}$, indeed produce a single filament
in which the spectrum broadens and the pulse
duration decreases upon propagation.
Experimental \cite{Stibenz:ol:31:274,Skupin:pre:74:056604} as well
as theoretical work
\cite{Kosareva:apb:91:35,Berge:prl:100:113902} indicated that
self-compression occurs in a limited parameter range, and small
deviations from these ideal conditions may either cause spatial
breakup into multiple filaments or temporal splitting of the pulse.
For evidencing self-compression, experiments currently employ
Brewster-windowed cells filled with noble gas at
pressures controlling the ratio of input power over critical.
Diagnostics are furthermore placed several tens of cm beyond the
sample, making the pulse undergo a last stage of atmospheric
propagation. A crucial point is then to preserve the main
characteristics of the compressed pulse when it exits the output
window. Several points concerning this issue, however, remain
unclear. First, the pulse entering the glass sample undergoes a
nonlinear index being three decades higher than in the gas.
This renders the robustness of the exiting optical
structure questionable as a light beam with several thousands of
critical powers is expected to immediately break up
by modulational instability
\cite{Bespalov:jetp:3:307,Campillo:apl:23:628, Berge:pd:176:181,
Fibich:oe:13:5897}. Second, the much stronger dispersion of glass
should severely alter the temporal compression. For
instance, a 1 mm thick sample with group-velocity dispersion (GVD)
coefficient $k^{(2)} \simeq 370$ fs$^2$/cm at 800 nm can double the
duration of a 8 fs transform-limited Gaussian pulse
\cite{Agrawal:NFO:01}.

In the following we will show that despite these apparent
sources of instability, filamentary propagation
holds a hidden self-healing mechanism for ultrashort pulses.
Comparable phenomena have already been observed in several physical
areas, e.g., in solid-state dynamics on atomic
scales \cite{Gerde:nat:413:285} and in plasma
physics under magnetic confinement \cite{Narihara:prl:87:135002},
explaining stabilization mechanisms in apparently unstable systems.
In optics, spatial self-restoration of Bessel beam profiles
\cite{Bouchal:oc:151:207} and of filament cores
\cite{Kolesik:ol:29:590,Dubietis:ol:29:2893} have also been
reported. In contrast to all these previous
observations, we report on simultaneous spatio-temporal
self-restoration of light bullets having experienced a perturbation
that should have immediately caused spatial and temporal decay of
the filament. In this Letter we identify the
changes in a self-compressed filament when it leaves a cell filled
with argon, propagates in a mm long silica window,
and finally travels through the atmosphere. We show that glass-dispersion
acts jointly with the Kerr response to widely stretch the pulse in
time. Unexpectedly, the filament eventually
recompresses in air to restore temporal profiles with few-cycle
durations. This self-healing mechanism has not an anecdotic character, as it
can be responsible for the remarkably short pulse durations seen in
many self-compression experiments.

Propagation equations model
the forward component of the pulse envelope, ${\cal E}(x,y,z,t)$, coupled
with the free electron density, $\rho(x,y,z,t)$
\cite{Berge:rpp:70:1633}
\begin{eqnarray}
\partial_z {\cal E} & = & \frac{i}{2k_0} {T}^{-1}\nabla_{\perp}^2
{\cal E}  + i{\cal D} {\cal E} - i\frac{k_0}{2n_0^2\rho_c} {T}^{-1}
\rho {\cal E} - \frac{\sigma}{2}\rho {\cal E} \label{1} \\
 & & - \frac{U_i W(I) \rho_{\rm nt}}{2 I}{\cal E} +
i\frac{\omega_0}{c} n_2 T \! \int \negmedspace {\cal R}(t-t')|{\cal
E}(t')|^2dt' {\cal E}, \nonumber \\
\partial_t \rho & = & W(I) \rho_{\rm nt} + \sigma\rho I/U_i -
\rho/\tau_{\rm rec},
\label{2}
\end{eqnarray}
where $I = |{\cal E}|^2$, $z$ is the propagation variable and $t$ the
retarded time in a reference frame moving with group velocity $1/k^{(1)}$
at laser mid-frequency $\omega_0$; $k(\omega)
= n(\omega) \omega/c$, $k_0=k(\omega_0)$, $n_0 = n(\omega_0)$ and $T =
1 + (i/\omega_0)
\partial_t$ \cite{Brabec:prl:78:3282}. ${\cal
D} = \sum_{n\ge2} (k^{(n)}/n!) (i \partial_t)^n$ is the dispersion
operator formally involving the derivatives $k^{(n)} \equiv
\partial^n k/\partial {\omega}^n|_{\omega_0}$. Linear dispersion
curves $n(\omega)$ are taken from Refs.
\cite{Dalgarno:procrsoca:259:424, Agrawal:NFO:01,Peck:josa:62:958}
for argon, silica and air, respectively. At 800~nm, $n_0$ takes the
value 1 for gases and 1.45 for glass. The leading GVD term in $\cal
D$ [$k^{(2)}$] is about three orders of magnitude larger in glass
than in the two gases; $\rho_c= 1.73 \times 10^{21}$ cm$^{-3}$
is the critical plasma density.
The nonlinear Kerr
response contains only an instantaneous contribution, ${\cal R}(t) =
\delta (t)$, when the pulse self-focuses in argon. It includes a
Raman-delayed contribution, ${\cal R}(t) = (1-f) \delta(t) + f
\theta(t)\frac{1+ \omega_R^2 \tau_R^2}{\omega_R \tau_R^2} \,
\mbox{e}^{-t/\tau_R} \sin{(\omega_R t)}$, with ratio $f$ in silica
\cite{Zozulya:prl:82:1430,Sprangle:pre:66:046418} and in air
\cite{Champeaux:pre:77:036406}. Plasma sources are driven by the PPT
ionization rate $W(I)$ \cite{Perelomov:spjetp:23:924} in argon and
air, and by the Keldysh rate for crystals in glass
\cite{Keldysh:spjetp:20:1307}. In the argon cell we
assume a uniform pressure of 0.5 bar, leading to a nonlinear Kerr
index $n_2 = 5 \times 10^{-20}$ cm$^2$/W at 800 nm. The Kerr index
for glass is $n_2 = 3.2 \times 10^{-16}$ cm$^2$/W
and $n_2 = 2.5 \times 10^{-19}$ cm$^2$/W in air at
atmospheric pressure \cite{Champeaux:pre:77:036406}. All other
parameter values used in this paper, including ionization potentials
$U_i$, avalanche cross-sections $\sigma$, neutral densities $\rho_{\rm nt}$,
and electron recombination
time for silica $\tau_{\rm rec}$,
can be found in Ref.~\cite{Skupin:pre:74:056604} (argon) and
Ref.~\cite{Berge:rpp:70:1633} (glass, air). Simulations have been
performed in full 3D ($\nabla_{\perp}^2 = \partial_x^2 +
\partial_y^2$) and radially-symmetric ($\nabla_{\perp}^2 = r^{-1}
\partial_r r
\partial_r,\,r = \sqrt{x^2 + y^2}$) geometries, yielding analogous results.
The starting pulse, being Gaussian, is focused by a lens of focal
length $f = 50$ cm into an argon cell with a 1.5 m
maximum length. Input waist at 1/e$^2$ and full width at
half-maximum (FWHM) duration are $w_0 = 500$ $\mu$m and $\tau_0
\simeq 30$ fs. The input power $P_{\rm in}$ is equal to one critical
power in argon, i.e., $P_{\rm cr} \simeq \lambda_0^2/2\pi n_0 n_2 =
20.4$ GW. The following analysis is divided into
three steps devoted to pulse propagation in argon, inside the silica
window, and propagation in air.

\begin{figure}
\includegraphics[width=\columnwidth]{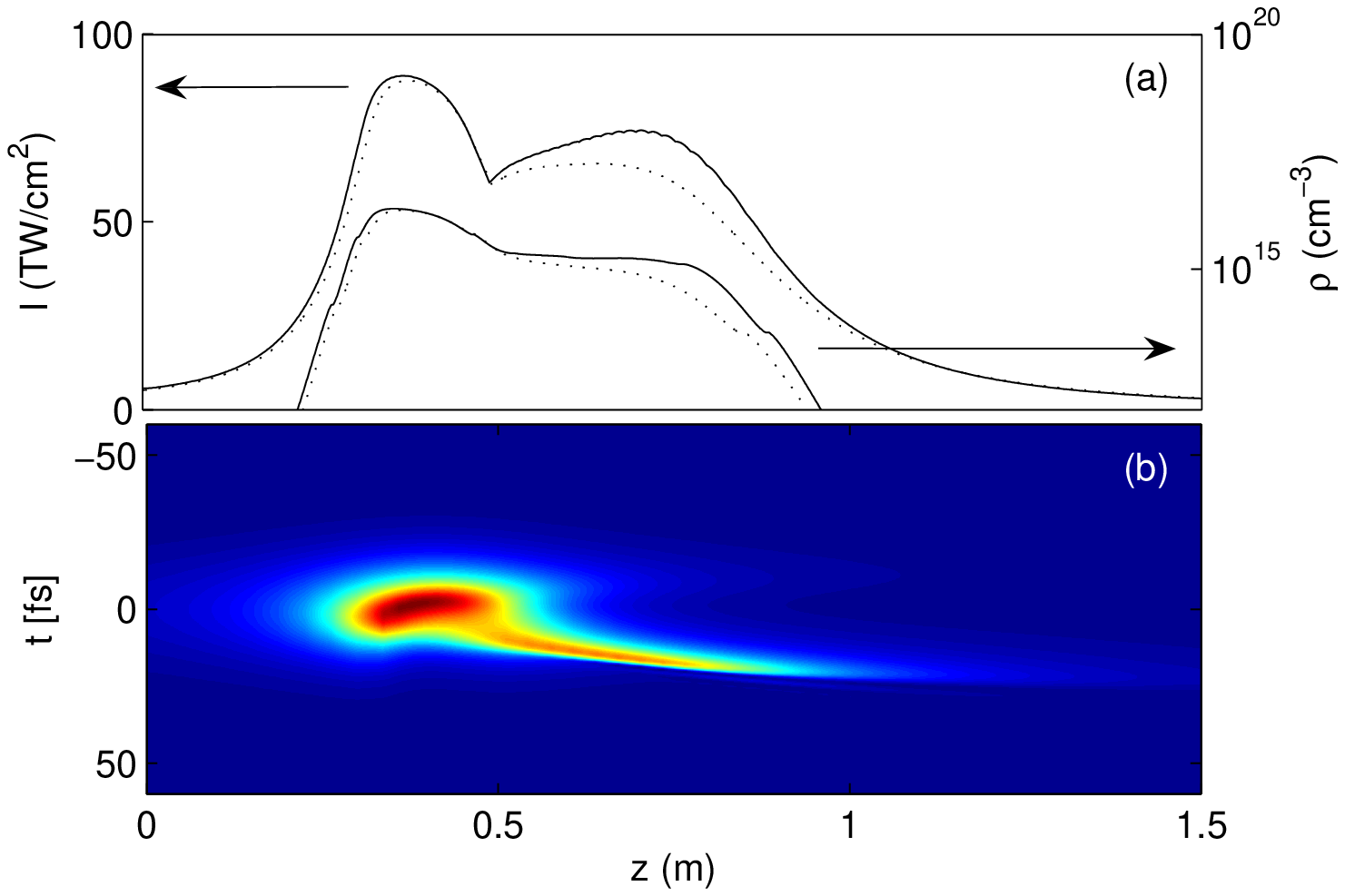}
\includegraphics[width=\columnwidth]{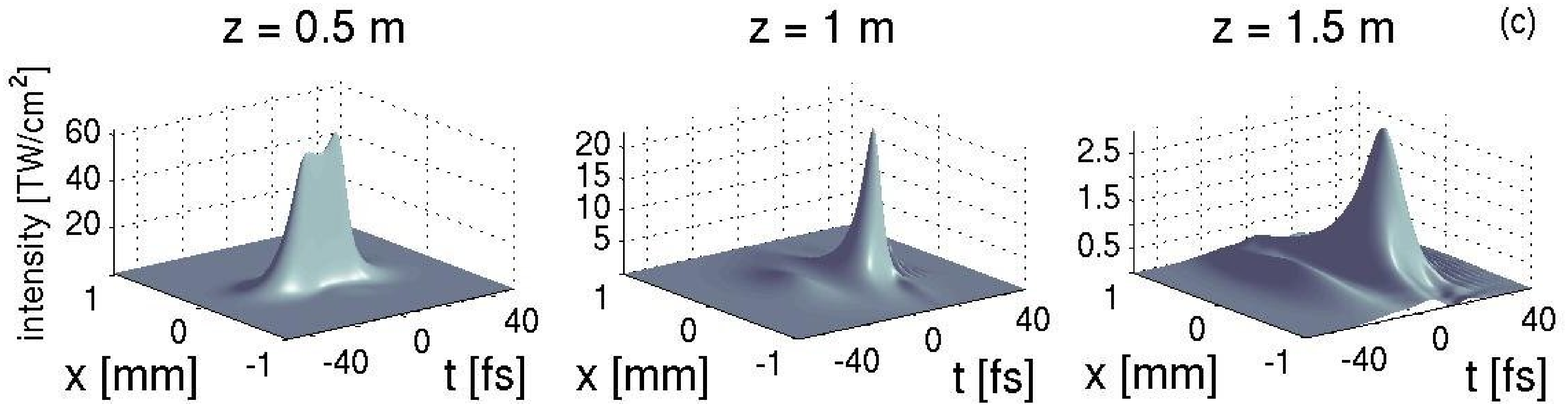}
\caption{(a) Peak intensity (measured on left axis) and electron density
(resp. right axis) for a 500
$\mu$m-waisted, 30 fs Gaussian pulse focused with $f = 50$
cm inside a 1.5 m long cell of argon computed in 3D geometry (solid curves)
and in radial
symmetry (dotted curves). 3D simulations involve a $5\,\%$ amplitude random
noise in the input.
(b) Temporal on-axis dynamics. (c) Intensity
profiles in the $(x,t)$ plane at the three positions selected for the exit
glass window.}
\label{fig1}
\end{figure}

{\it 1. Filamentation in argon}. With one critical power, the pulse
develops a light bullet dynamics \cite{Berge:prl:100:113902},
characterized by a primary focusing sequence, followed by a second,
limited refocusing [Fig. \ref{fig1}(a)]. Minimum
FWHM duration is attained when the plasma
relaxes, and the filament is compressed to about 6
fs at $z \simeq 0.7$ m [Fig. \ref{fig1}(b)]. The three subplots of
Fig. \ref{fig1}(c) show intensity distributions at
distances $z = 0.5,\,1$ and $1.5$ m where the on-axis fluence
(${\cal F} \equiv \int |{\cal E}|^2 dt$) reaches 1.7 J/cm$^2$, 0.2
J/cm$^2$, and 0.05 J/cm$^2$, respectively. These distances
define the effective length of the argon cell and
the position of the exit glass window.

\begin{figure}
\includegraphics[width=\columnwidth]{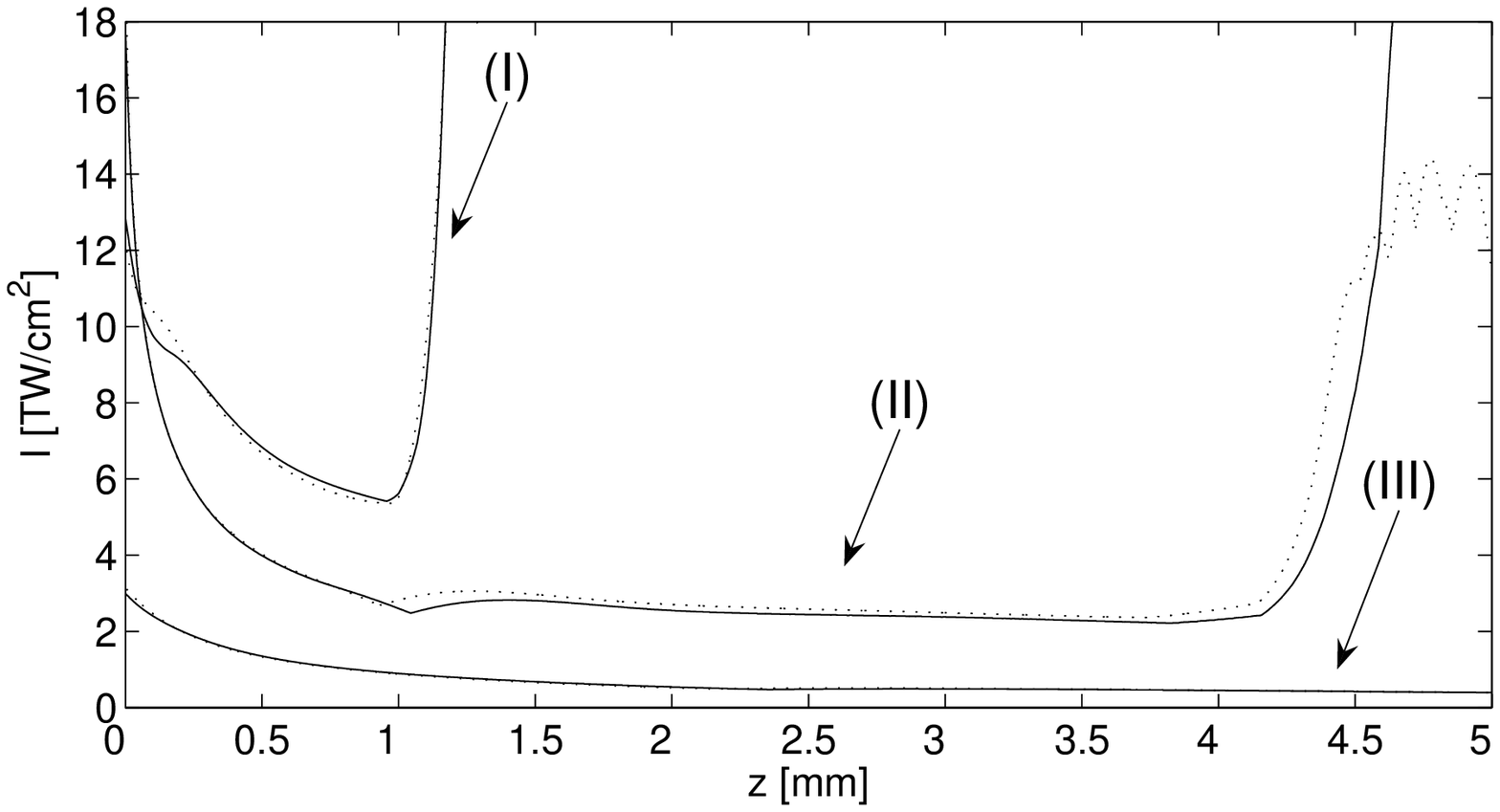}
\includegraphics[width=\columnwidth]{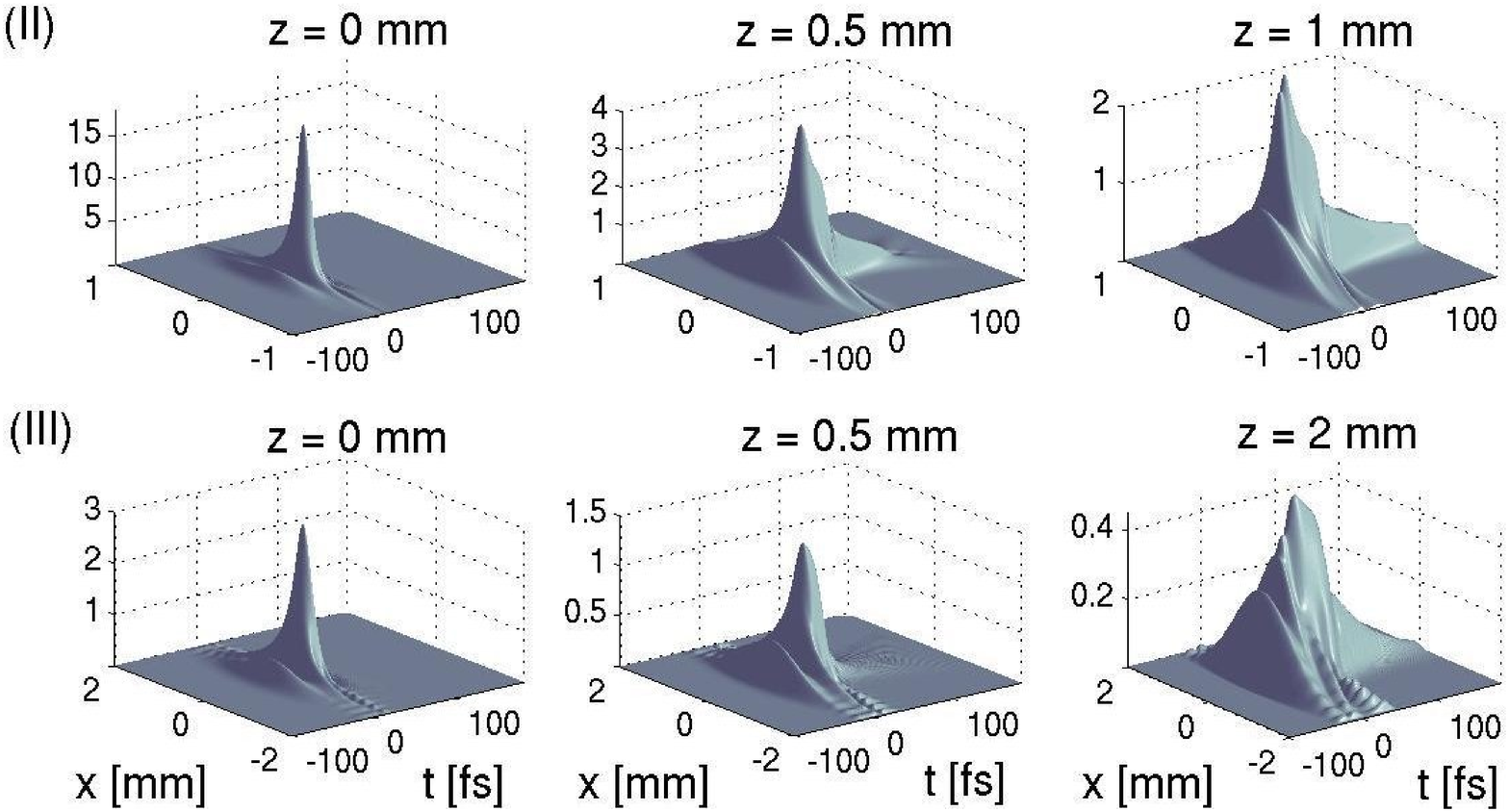}
\caption{(Top) Peak intensity of the filament exiting the argon cell
at (I) $z = 0.5$ m, (II) $z = 1$ m and (III) $z =
1.5$ m, and propagating into the silica window over mm distances.
(Middle and bottom) Rows (II) and (III)
detail the $(x,t)$ evolution of the pulse profiles II and III inside
the exit window.}
\label{fig2}
\end{figure}

{\it 2. Crossing the silica window}.  At high enough fluences $>
0.1$ J/cm$^2$, only a fraction of the forward pulse can be
transmitted through a silica surface. According to
\cite{Penano:pre:72:036412}, only 20 $\%$ of the
forward pulse is expected to transmit the glass
surface at $z = 0.5$ m, whereas this fraction
increases to 90 $\%$ and 100 $\%$ for $z = 1$ and 1.5 m,
respectively. Conditions beyond $z=1$ m ($I < 2
\times 10^{13}$ W/cm$^2$) therefore avoid damage of silica
\cite{Rayner:oe:13:3208}. Figure \ref{fig2}(a) summarizes our
results. Output profiles of Fig. \ref{fig1}(c) multiplied by the
previous transmission rates are used as initial conditions of Eqs.
(\ref{1})-(\ref{2}) solved for silica. 3D and radially-symmetric
simulation results almost superimpose; hence, no severe spatial
distortions affect the intensity profiles in the $(x,y)$ plane. The
pulse captured at $z = 0.5$ m (pulse I) begins to diffract over 1
mm, but it refocuses and collapses just afterwards. Although
experimentally irrelevant, this extreme configuration already points
out the importance of the first few
mm of propagation dominated by pulse dispersion. Triggered by beam
fluences $>1.5$ J/cm$^2$ in argon, the surface
plasma created by pulse I should, however, act as a filter
transmitting only the leading edge of the pulse
\cite{Feit:apa:79:1657}, thereby rendering the
underlying assumption of a fixed temporal profile invalid.
For pulses II and III having much lower fluences, assumption of a fixed profile 
still holds. In the following, we thus concentrate on these last two configurations
only.

When the window is positioned at $z = 1$ m (pulse II), again
dispersion first prevails, but does not prevent
beam collapse over 5 mm long paths. The input beam has a waist of
240 $\mu$m and conveys time slices with a maximum
power of $\sim 7100\,P_{\rm cr}$, where $P_{\rm cr}^G$ = 2.19 MW
denotes the critical power in glass. In contrast, the low-intensity
pulse III disperses rather slowly over the same distance and beyond.
The maximum input power, however, is still about
8000 $P_{\rm cr}$, as the beam waist entering the sample has been
enlarged to $\sim 0.6$ mm. Rows (II) and (III) of Fig. \ref{fig2}
depict the evolution of pulses II and III in the $(x,t)$ plane.
Pulse II enters the dielectric with time extent of $\sim 8$ fs.
After $0.5$ mm in silica, this pulse decays into a broader structure
of $\sim 33$ fs, forming a pedestal that extends
deep into the trailing region. After 1 mm, its
duration increases to $50$ fs. Pulse III first
exhibits a duration of 13 fs, which then broadens
to 28 fs over 0.5 mm in glass and more afterwards.

It is important to note the complete absence of
multifilamentation (MF) in our simulations. Estimates from
modulational instability theory \cite{Campillo:apl:23:628}, however,
predict an onset for MF and subsequent collapse upon short
distances, $\Delta z_{\rm MF} \simeq n_0 P_{\rm cr}/\lambda_0 I_{\rm
max} \sim 0.1$ mm, for peak intensities $I_0 \geq 3$ TW/cm$^2$
impinging on the glass surface. In fact, as GVD
rapidly damps the peak intensity, the amplification
of periodic spatial modulations leading to MF is practically
prevented. Here the plasma density plays a negligible role in the
nonlinear index, as $\Delta n_{\rm NL} \simeq n_2 I_{\rm max} -
\rho_{\rm max}/2 n_0 \rho_c \sim 1-5 \times 10^{-3}$. We also
numerically checked that high-order dispersion, pulse steepening and Raman 
scattering are of minor influence.

Besides, it turns out that both GVD and the Kerr response are
responsible for the strong increase of the pulse
duration that exceeds predictions by the well-known linear formula
for Gaussian pulses, $\tau_{\rm lin}(z) \simeq \tau_0 \sqrt{1 + 16
(\ln{2})^2 (k^{(2)} \Delta z/\tau_0^2)^2}$ \cite{Agrawal:NFO:01}. Denoting the
spatial and temporal radii of Gaussian pulses by $R(z)$ and $T(z) =
\tau(z)/2 \sqrt{\ln{2}}$, respectively, a two-scaled variational
principle applied to the reduced equation $i \partial_z \psi +
\nabla_{\perp}^2 \psi - \delta
\partial_t^2 \psi + |\psi|^2 \psi = 0$ can help us understand these
peculiar dynamics through the system \cite{Skupin:pd:220:14}
\begin{equation}
\label{VA}
\frac{1}{4} R^3 R_{zz} = 1 - \frac{p}{2T}\,\,\,; \,\,\,
\frac{1}{4} T^3 T_{zz} = \delta ( \delta + \frac{T
p}{2R^2}).
\end{equation}
In Eq.\ (\ref{VA}), $p = P_{\rm in}/P_{\rm cr}$ and $\delta \equiv 2
\pi n_0 R_{\rm fil}^2 k^{(2)}/\lambda_0 T_{\rm fil}^2$ is the
normalized GVD coefficient for the input filament with size $R_{\rm
fil}$ and temporal radius $T_{\rm fil}$. $\delta$ takes very large
values $> 200$ when using durations and waists of pulses II and III.
Since the growth of $T(z)$ is driven by $\delta^2$ (dispersion) and
$\delta \times p$ (Kerr response), the beam
collapse characterized by the vanishing $R(z) \rightarrow 0$ at
finite $z$ can efficiently be delayed in the limits $\delta,\,p \gg
1$. Indeed, the more $T(z)$ increases, the less $R(z)$ decreases.
Moreover, Eq.\ (\ref{VA}) illustrates that
both GVD and cubic nonlinearity
contribute to enlarge the pulse duration.
Typically, FWHM durations of $\sim 25$ fs are attained over 0.5 mm
by the initially 8 fs long pulse II, whereas linear
dispersion alone allows a temporal broadening limited to $\sim 10$
fs. For this pulse, broadening is more than doubled by the Kerr
response. Hence, GVD disperses the pulse while at high powers the
Kerr term stretches it in time. Combining both effects
causes significant pulse broadening.

\begin{figure}
\includegraphics[width=\columnwidth]{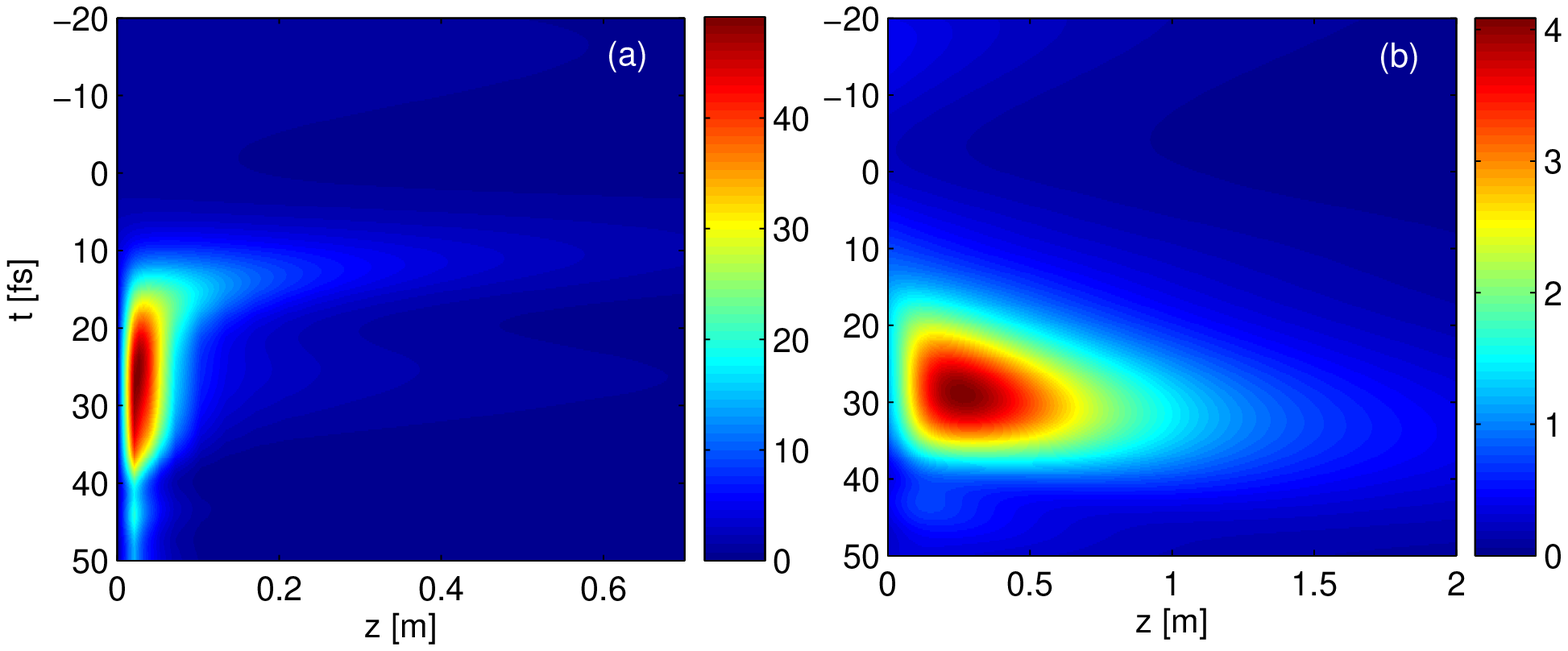}
\includegraphics[width=\columnwidth]{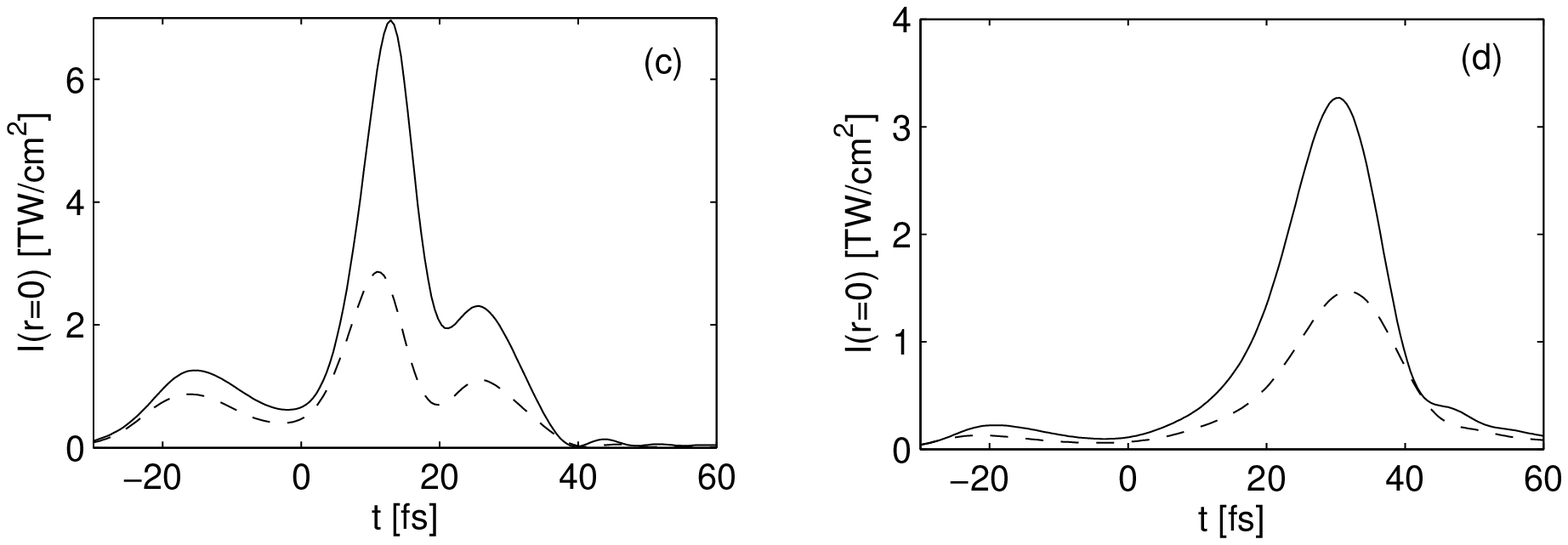}
\caption{On-axis temporal evolution of (a) pulse II and (b) pulse III in the
atmosphere, after exiting a $0.5$ mm
thick window. (c) Temporal profiles compressed in air at the distances
$z = 0.25$ m (solid line) and
0.5 m (dash-dotted line) for pulse II. (d) Same for pulse III at
$z = 0.5$ m (solid line) and $z = 1$ m (dashed line).}
\label{fig3}
\end{figure}

{\it 3. Air propagation.} Because diagnostics are usually positioned
tens of cm away from the cell, the beam finally
propagates in atmosphere. Figures \ref{fig3}(a,b)
show the temporal evolution of pulses II and III as they exit a 0.5
mm thick window and are fully transmitted in air. These two
configurations are characteristic of filaments diagnosed either at
moderately high intensities (II) or close to the end of the
diffraction stage (III). Because these pulses contain time slices
with maximum powers of $\sim 3.5$ $P_{\rm cr}$ in
air, they eventually refocus and thus recompress,
which is the core of the self-healing process.
Pulse II attains intensities less than 50 TW/cm$^2$ and triggers a
minor plasma response. Pulse III involves much weaker intensities.
The key-player of this process is the self-focusing mechanism alone,
which amplifies the time slice with the largest power to the
detriment of the others \cite{Berge:rpp:70:1633}. Collapse is not
arrested by plasma generation, but instead by Raman-scattering. The
Raman-delayed response indeed doubles the effective power for
self-focusing and delocalizes the pulse in time. At powers close to
critical, this is sufficient to arrest the beam collapse. 
Figures \ref{fig3}(c,d) detail temporal profiles
with FWHM durations of $10.3$ fs at $z = 0.25$ m and $10.8$ fs at $z
= 0.5$ m for pulse II, and of $16.6$ fs and $18.6$ fs at 0.5 m and 1
m for pulse III. These durations are relatively stable since they
hold over several tens of cm. They remain in the order of magnitude
of those reported in the experiments \cite{Stibenz:ol:31:274}.

\begin{figure}
\includegraphics[width=\columnwidth]{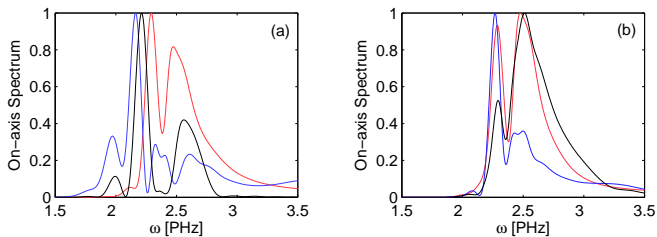}
\caption{On-axis spectra for (a) pulse II and (b) pulse III. Red curves refer
to the dynamics after compression in argon. Blue curves show that
in the window at $z = 0.5$ mm. Black curves refer to air propagation
with data collected at $z = 0.5$ m.}
\label{fig4}
\end{figure}

Figure \ref{fig4} summarizes the history of on-axis spectra around
the pump frequency ($\omega_0 = 2.35$ PHz) along the propagation
steps 1 to 3. Figure \ref{fig4}(a) describes pulse II, passing through
the glass window and transmitted to air. The first focusing event in
argon creates red- and blue-shifted spectral wings caused by
self-phase modulation (red curve). Next, GVD drives
the propagation in the dielectric, which lowers the saturation
intensities and weakens self-phase modulation, such
that the blueshifted wing drops out (blue curve).
Figure \ref{fig4}(b) provides the same pieces of
information for pulse III. At weaker intensities,
the spectral evolution follows that of pulse II, although
the properties expected from linear dispersion
become more important and plasma turns off. Hence,
on-axis spectra remain closer to those exiting argon.

It is worth emphasizing that other compressor schemes based, e.g., on an Ar-glass-Ar interface, 
were also numerically simulated (not shown). 
Increasing the pressure to 1 bar in the second argon cell guarantees a power above critical 
for pulse II, which then refocuses and follows a temporal evolution similar to 
that of Fig. \ref{fig3}(a). This underlines the generality of the self-healing mechanism. 

In conclusion, numerical simulations cleared up the dynamics of
self-compressed filaments when they cross a gas-glass-gas interface.
After they are produced
in a noble gas, they propagate over a few mm inside a silica window
and run tens of cm in air before being measured. In the window, the interplay between 
dispersion and Kerr self-focusing inhibits
multifilamentation and significantly increases
pulse durations. Despite this large stretching in time,
durations in the 10 fs range restore themselves upon
further nonlinear propagation in air. This new self-healing mechanism 
acts both in space and time. Unlike
previous scenarios where the perturbation is caused
by loss mechanisms, such as raindrops, apertures or
wire grids, the self-restoration is here induced by a discontinuity of
the nonlinear properties inherent to the materials.

\end{document}